\documentclass[12pt,preprint]{aastex}

\def\meth {CH$_3$OH~}
\def\hho  {H$_2$O~}
\def\hii  {\ion{H}{2}~}
\def\SgrA {Sgr~A*}
\def\wsrc {W51~Main/South}
\def\msrc {W51~Main}
\def\ssrc {W51~South}
\def\nsrc {W51~North}
\def\isrc {W51~IRS~2}

\def\deg  {\ifmmode {^\circ}\else {$^\circ$}\fi}
\def\porm {\ifmmode {\pm}\else {$\pm$}\fi}
\def\Vlsr {\ifmmode {V_{\rm LSR}} \else {$V_{\rm LSR}$}\fi}
\def\Ro   {\ifmmode {R_0} \else {$R_0$} \fi}
\def\To   {\ifmmode {\Theta_0} \else {$\Theta_0$} \fi}

\def\d    {\ifmmode {{\rlap{.}}^\circ}\else {${\rlap{.}}^\circ$}\fi}
\def\s    {\ifmmode {{\rlap{.}}^s}\else {${\rlap{.}}^s$}\fi}
\def\as   {\ifmmode {{\rlap{.}}{''}}\else {${\rlap{.}}{''}$}\fi}

\def\kms  {km~s$^{-1}$}
\def\masy {mas~yr$^{-1}$}
\def\jyb {Jy~beam$^{-1}$}

\def\etal {et al.}
\def\eg   {e.g.,~}
\def\ie   {i.e.,~}

\slugcomment{Draft version: August 4, 2010}
\shorttitle{TRIGONOMETRIC PARALLAX OF W51 MAIN/SOUTH} 
\shortauthors{Sato \etal}

\begin{document}

\title{TRIGONOMETRIC PARALLAX OF W51 MAIN/SOUTH}

\author{M. Sato\altaffilmark{1,2,3}, M. J. Reid\altaffilmark{2}, 
 A. Brunthaler\altaffilmark{4}, and K. M. Menten\altaffilmark{4}}
\altaffiltext{1}{Department of Astronomy,
 Graduate School of Science,
 The University of Tokyo, Tokyo 113 0033, Japan}
\altaffiltext{2}{Harvard-Smithsonian Center for
 Astrophysics, 60 Garden Street, Cambridge, MA 02138, USA}
\altaffiltext{3}{VERA Project, National Astronomical Observatory,
 Tokyo 181 8588, Japan}
\altaffiltext{4}{Max-Planck-Institut f\"ur Radioastronomie,
 Auf dem H\"ugel 69, 53121 Bonn, Germany}

\begin{abstract}
We report measurement of the trigonometric parallax of \wsrc\ using
 the Very Long Baseline Array (VLBA).
We measure a value of 0.185\porm 0.010~mas, corresponding to
 a distance of 5.41$^{+0.31}_{\,-0.28}$~kpc.
\wsrc\ is a well-known massive star-forming region
 near the tangent point of the Sagittarius spiral arm
 of the Milky Way.
Our distance to W51 yields an estimate of the 
 distance to the Galactic center of
 $\Ro=8.3\pm 0.46$ (statistical) $\pm 1.0$ (systematic) kpc by simple geometry.
Combining the parallax and proper motion measurements for W51, we
 obtained the full-space motion of this massive star forming region.
We find W51 is in a nearly circular orbit about the Galactic center.
The H$_2$O masers used for our parallax measurements trace
 four powerful bipolar outflows within a 0.4 pc size region, some
 of which are associated with dusty molecular hot cores and/or hyper- or
 ultra-compact \hii regions.
\end{abstract}


\section{Introduction}
We are using the National Radio Astronomy
 Observatory's\footnote[5]{The National Radio Astronomy Observatory
 is a facility of the National Science Foundation operated under
 cooperative agreement by Associated Universities, Inc.}
 Very Long Baseline Array (VLBA), the Japanese VERA
 (VLBI Exploration of Radio Astrometry) array
 and the European VLBI Network (EVN)
 to measure trigonometric parallaxes of
 maser sources in massive star-forming regions that define the
 spiral arms of the Milky Way galaxy 
 \citep{Xu:06, Xu:09, Hachisuka:06, Hachisuka:09, Honma:07,
 Hirota:07, Menten:07, Sato:08, Sato:10G, Choi:08, Kim:08,
 Reid:09I, Reid:09VI, Reid:09R0, Moscadelli:09, Zhang:09,
 Brunthaler:09, Moellenbrock:09, Sanna:09, Rygl:10, Oh:10}.
The primary goal of these efforts is to delineate the spiral
 structure of the Milky Way by direct and accurate distance
 determination.
Our current view of Galactic structure relies
 mostly on uncertain kinematic distances, which tend to
 overestimate the source distances, sometimes over a factor of 2
 \citep[\eg][]{Reid:09VI, Sato:10G}.
Therefore, a precise map of the Galaxy based on trigonometric
 parallaxes of massive star-forming regions will significantly
 refine our knowledge of Galactic structure.
In addition to parallax measurements, we also obtain source proper
 motions.
Combining the proper motion and distance of the source with its
 radial velocity yields the full three-dimensional (3D) space motion of
 the source, allowing detailed study of
 the kinematics of the Galaxy.

In this paper, we report parallax measurements of \wsrc, a
 well-studied massive star-forming region that hosts some of the
 strongest \hho maser sources in the Galaxy.
\wsrc\ is associated with a giant \hii region W51~A
 \citep{Figueredo:08}, where ongoing formation of massive stars has
 been reported \citep[\eg][]{Okumura:00, Clark:09}.
A reliable distance to W51 is important for
 the physics of massive star formation, since masses, luminosities and
 ages of young stellar objects depend sensitively on distance.
Also, W51 has special importance for the study of Galactic structure,
 owing to its location very close to the tangency of the Sagittarius
 spiral arm, hence yielding a good estimate of the distance to the
 Galactic center by simple geometry.  

\section{Observations and Data Analysis}

\subsection{Observations} 
We used the NRAO VLBA to observe the 22 GHz \hho maser source in
 the massive star-forming region \wsrc. 
The observations were conducted under VLBA program BR134,
 which included observations of four star-forming regions
 thought to be in the Sagittarius spiral arm of the
 Milky Way.
Observations of the other sources will be
 reported in a forthcoming paper.

Prior to the VLBA parallax observations, we used
 the NRAO Very Large Array (VLA; program
 AR677) in the most extended A configuration
 on 2008 September 28 to find background extragalactic
 sources as position references near the target \hho maser sources.
Following the methods described by \citet{Reid:09I}, we selected
 unresolved sources in the NRAO VLA Sky Survey (NVSS) catalog
 within $\approx2\deg$ of each maser target and observed both
 background candidates and target masers at 8 and 22 GHz.
For \wsrc, we found one new background source J1922+1504 for VLBA
 observations in addition to two known calibrators from the VLBA
 Calibrator Survey-1 \citep[VCS1; ][]{Beasley:02}, J1922+1530 and
 J1924+1540.
However, at 0.3 mas resolution of the VLBA, we found the new source
 J1922+1504 displayed a resolved structure
 and was not useful for parallax measurements
 (see \S\ref{sect:parallax}).  
We also obtained an accurate position of
 the \hho maser source in \wsrc\ to observe with the VLBA.

The VLBA observations of \wsrc\ were scheduled on 2008 October 22,
 2009 April 27 and 30, and 2009 October 20, as listed in Table
 \ref{table:obs}.
These dates well sample the peaks of the sinusoidal parallax
 signature in right ascension, maximizing the sensitivity of
 parallax detection and ensuring that the parallax and proper
 motion are essentially uncorrelated.
We chose four optimal epochs to sample the parallax peaks
 in right ascension only, instead of sampling the peaks both
 in right ascension and declination over six or more epochs,
 since the parallax signature of W51 in right ascension
 is about twice as large as in declination.
The two middle epochs were chosen close to each other 
 and near one parallax peak (which we call a ``1--2--1'' schedule).
The two middle epochs are needed for symmetry and to obtain
 similar accuracy for the parallax maximum and minimum.
This yields zero correlation among the parallax and
 proper motion parameters.
Compared to a ``1--1--1--1'' schedule over 1.5 cycles (years) of
 the parallax signature, this ``1--2--1'' schedule also reduces
 the time span of observations to one year,
 reducing sensitivity to maser spot lifetimes. 
 
Our 22 GHz observations were set up in the same manner as
 described by \citet{Reid:09I} for 12 GHz \meth maser observations.
For atmospheric calibration, four ``geodetic'' blocks of ICRF 
 (the International Celestial Reference Frame) quasars
 \citep{Ma:98, Fey:04} with positions accurate to $<1$~mas were
 observed at different antenna elevations across the sky in
 left-circular polarization with eight 4 MHz bands that spanned 450
 MHz of bandwidth between 22.02 and 22.47 GHz and uniformly sampled
 all frequency differences.
Using a four-block setup allows the middle block of observing time to be
 available for phase-referencing rapid-switching scans between the
 target maser and background continuum sources when the source
 elevation was the highest at most stations.
We measured multi-band delays and
 fringe rates, mostly due to un-modeled atmospheric propagation delays,
 to determine zenith delay errors as a function of
 time for each antenna.

Table \ref{table:pos} lists source information for the target \hho
 maser emission in \wsrc\ and background continuum sources.
We placed three rapid-switching (phase-referencing) blocks between the
 geodetic blocks and switched every 30~s 
 between the target maser and one of the three background
 continuum sources.
For rapid-switching blocks, we employed four adjacent frequency
 bands of 8 MHz bandwidth in both right- and left-circular
 polarizations.  
The four bands were centered at local standard of rest (LSR)
 velocities, \Vlsr, of 276.0, 168.0, 60.0, and $-$48.0 \kms,
 covering the maser velocity extent
 reported in previous studies
 \citep{Genzel:81, Imai:02};
 \hho maser emission was detected in the second, third, and
 fourth bands.

The data correlation was performed at the VLBA correlation facility
 in Socorro, NM.
The data from each antenna pair were cross-correlated with an
 integration time of 0.92~s.
The expected time-averaging effect in wide-field mapping
 from this integration time, for a beam size of
 $\approx 0.3$~mas, is
 5\% loss in peak amplitude at an angular distance of
 $\Delta\Theta \approx 2''$ from the correlation center
 \citep[\eg][]{Thompson:01}. 
Each of the frequency bands was split into 256
 spectral channels, yielding frequency and velocity resolutions of
 31.25~kHz and 0.42~km~s$^{-1}$, respectively, for a rest frequency
 of 22.235080~GHz \citep{Pickett:98} for the
 H$_2$O 6$_{16}\rightarrow$5$_{23}$ transition.

\begin{figure}
\epsscale{1.} 
\plotone{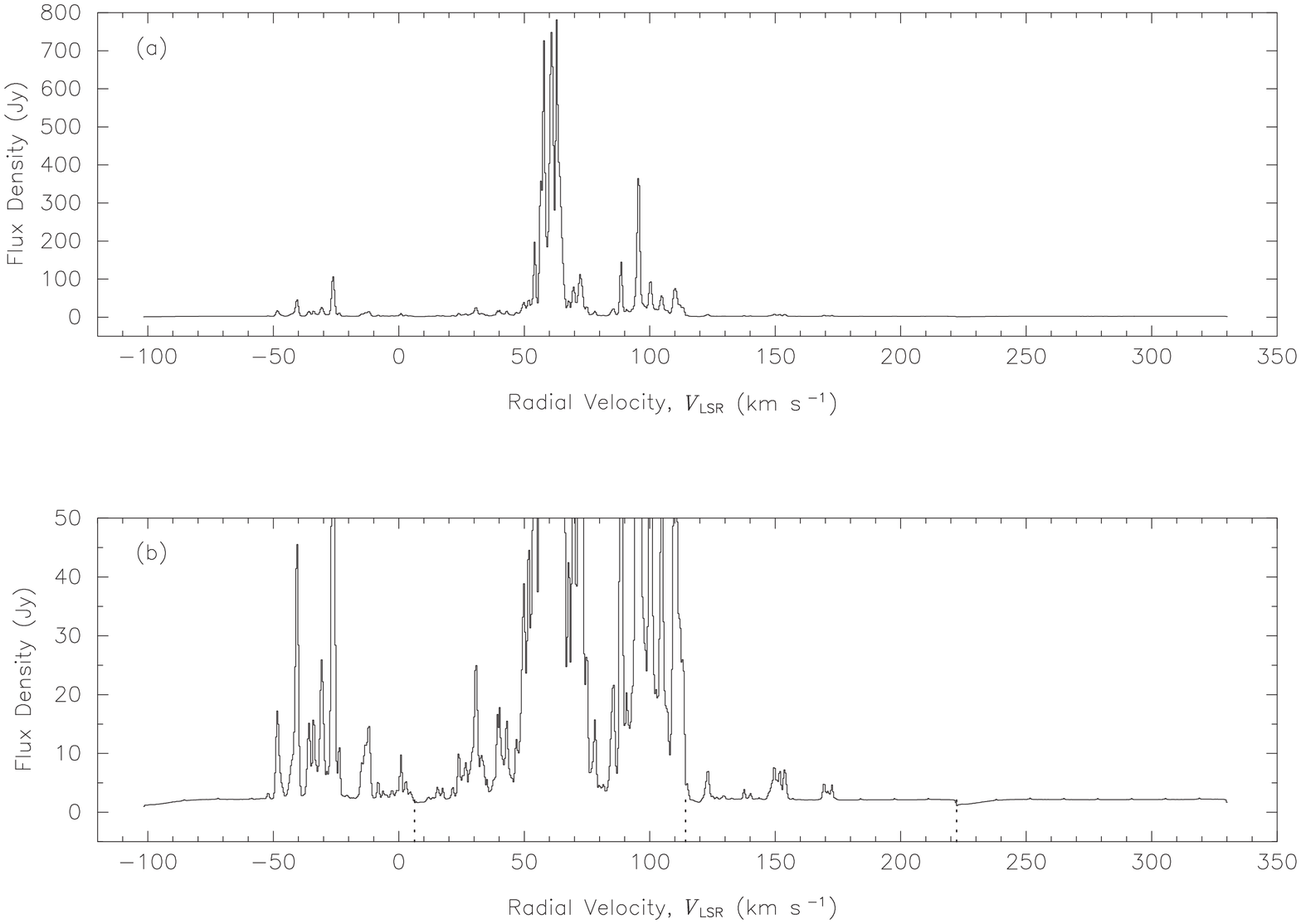} 
\caption{22 GHz \hho maser spectrum of \wsrc\ over all four
 8 MHz frequency bands from the first-epoch data.  Upper panel:
 on full scale.  Lower panel: blowup for weaker emission.
 Dotted vertical lines indicate the boundaries of four 8 MHz
 frequency bands observed.}
\label{fig:spect}
\end{figure}

Figure \ref{fig:spect} shows the 22 GHz \hho maser spectrum of
 \wsrc\ over all four 8 MHz frequency bands from the first-epoch's
 data produced by scalar averaging of the data over the whole
 observing time at all antennas except Hancock (flagged) and
 Saint Croix (no data).

\subsection{Calibration} 

We calibrated the correlated data using the NRAO Astronomical Image
 Processing System \citep[AIPS; ][]{Greisen:03} in the manner
 described in four steps by \citet{Reid:09I}.
We performed ``manual phase calibration'' using J2253+1608 data to
 remove instrumental delay and phase and delay offsets among the four
 frequency bands.
We then used one strong maser channel as the phase reference and
 subtracted reference phases of this channel from all data,
 including all maser channels and all background sources,
 by interpolating phase solutions in time. 
Since the position of J1924+1540 was known to 0.3 mas accuracy (and
 that of J1922+1530 to 0.8 mas accuracy) \citep{Beasley:02},
 the apparent map offset of J1924+1540 when phase referenced to
 a maser spot is due to a positional offset of the reference maser
 spot from the correlation center.
We corrected for this positional offset by shifting the phase center
 of the reference maser spot with the AIPS task CLCOR and
 re-fitted the phase reference data.
For the first through third epochs, we used the maser spot at
 $\Vlsr=62.9$ \kms\ in \ssrc\
 (feature 32 in Table \ref{table:intmot})
 as phase reference, after shifting the phase center by
 $(\Delta \alpha \cos \delta, \Delta \delta)
 = (0\as 8816, --1\as 1012)$
 to the reference spot position. 
Since this spot became significantly weaker (from 754 to 21 \jyb)
 and stronger emission appeared elsewhere in the same spectral channel
 as our observations progressed, we could no longer use this spot
 as a phase reference at the fourth epoch and used another spot at
 $\Vlsr=41.5$ \kms\ in \ssrc\
 (feature 26 in Table \ref{table:intmot}),
 after shifting the phase center by
 $(\Delta \alpha \cos \delta, \Delta \delta) =
 (1\as 2451, --1\as 1803)$ to its position.

We flagged jumps greater than 60\deg\ in the reference phase
 in adjacent maser scans and all data between these scans. 
We also flagged all data from antennas 
 that were not of good enough quality to give opacity correction
 and produced discrepant amplitude results.
Table \ref{table:obs} lists all stations that were either not used
 during the observation or flagged in the calibration.

\subsection{Finding Maser Spots}

Since maser spots are spread over a large angle on the sky, 
 we first made large-field image cubes
 involving all spectral channels for each 8 MHz frequency
 band from the first-epoch data covering a region of
 $\approx20'' \times 20''$ around the correlation phase center (\ie
 the phase center position was shifted back with CLCOR from the
 reference maser position).  
For the large-field imaging, we used only shorter baselines among the
 inner five VLBA antennas:  Fort Davis, Kitt Peak, Los Alamos, Owens
 Valley, and Pie Town.
Robust weighting was adopted for
 weighting of the visibility data \citep{Briggs:99}.
Each image cube was 4096 pixels by 4096 pixels over 256 channels
 with pixel size of 1
 mas, yielding a field of view of $\approx4'' \times 4''$.
We mapped 27 sub-fields per channel: a $5\times5$ grid spaced by
$4\as 096$ plus
2 sub-fields offset north and south of the center by $12\as 288$.
In order to remove the effects of sidelobes of strong maser
 features in the same spectral channel, all 27 sub-fields per
 channel were simultaneously CLEANed using the task
 IMAGR. 

After the $20'' \times 20''$ large-field imaging, we used
 Gaussian fitting with the AIPS task SAD (search and destroy)
 to detect maser spots.
Our criteria for maser spot detection required
 that each detected spot had a peak intensity greater than
 seven times the rms noise level ($7\sigma$) over
 two or more spectral channels within $\pm 2$ channels (\ie
 within $\Delta \Vlsr=$0.84 \kms) at the same position within
 $\pm 1$ mas in each coordinate.
From the first-epoch's data, we detected 1362 spots in 280
 unique ``features'' (\ie spots at the same position but in
 adjacent velocity channels).

\subsection{Mapping Maser Spot Distributions}

After the maser spot detection from large-field imaging, we
 imaged a small field of 1024 pixels by 1024 pixels with pixel
 size 0.03 mas ($\approx 1/10$ of the FWHM of the beam) centered
 at each spot position over detected channels, which yielded a
 field of view of $\approx$ 30 mas $\times$ 30 mas. 
If two or more spots were in overlapping fields, we used
 one field of 2048 pixels by 2048 pixels 
 centered at the average position of the spots.
All fields in the same channel were simultaneously CLEANed to
 remove sidelobes.
We shifted phase centers using CLCOR to
 each field before mapping the field.
 (For a technical detail, since we could only shift the phase
 center to one position at a time, we shifted the phase center
 using CLCOR to one of the fields that were simultaneously
 CLEANed, then kept only the image of the phase-centered field
 by discarding the other field images,  
 and repeated shifting and imaging for all fields.)
We then mapped each field at all four epochs in exactly the
 same manner, except we doubled the field size for the
 fourth epoch to allow for a large proper motion of the spot
 and/or small residual position offsets of the field centers
 due to the change of the reference maser spot at the fourth
 epoch.

After obtaining all four-epoch maps around each maser spot, we
 measured positions with the AIPS task SAD.
We calculated spot positions relative to the reference maser
 spot at $\Vlsr=62.9$ \kms\ in \ssrc\ (feature 32 in Table
 \ref{table:intmot}), by subtracting the position of this spot
 from all spot positions.  
We then fitted a linear (relative) proper motion for each spot
 using the position versus time measurements.
We identified
 a maser spot to be the ``same'' spot at different epochs if (1) the
 spot was detected at all four epochs as the strongest spot in the
 same imaging field, with an intensity greater than 0.5 \jyb\ and
 also greater than 10 times the rms noise of the map; 
 (2) the fitted relative proper motion was
 smaller than 10 \masy\ ($\approx250$~\kms\ at
 a distance of 5.4 kpc); and (3) the uncertainty for the fitted
 relative proper motion was smaller than $0.5$ \masy.

We grouped the identified maser spots into a maser feature (\ie a
 group of spots with similar position and velocity) if 
 (1) the positions were within 1~mas in each coordinate; (2) the
 radial velocities, \Vlsr, agreed within 7 \kms\ (which allows
 for maser emission from different hyperfine split components;
 \eg \citealt{Walker:84}); and
 (3) the fitted relative proper motions agreed within 0.3 \masy.
Table \ref{table:intmot} lists all 37 features
 with proper motions detected from all four epochs.
We estimated the radial velocity for each feature as the 
intensity-weighted mean of the radial velocities of the spots.
The position and proper motion of each feature relative to
 the reference spot were estimated as error-weighted mean of the
 values for the included spots.
Figure \ref{fig:intmot} shows the proper motions of the 
 detected features relative to the calculated mean motion
 of all regions (see Sections \ref{sect:intmot} and \ref{sect:mean}).

\section{Results}

\subsection{Parallax}\label{sect:parallax}

\begin{figure}
\epsscale{1.} 
\plotone{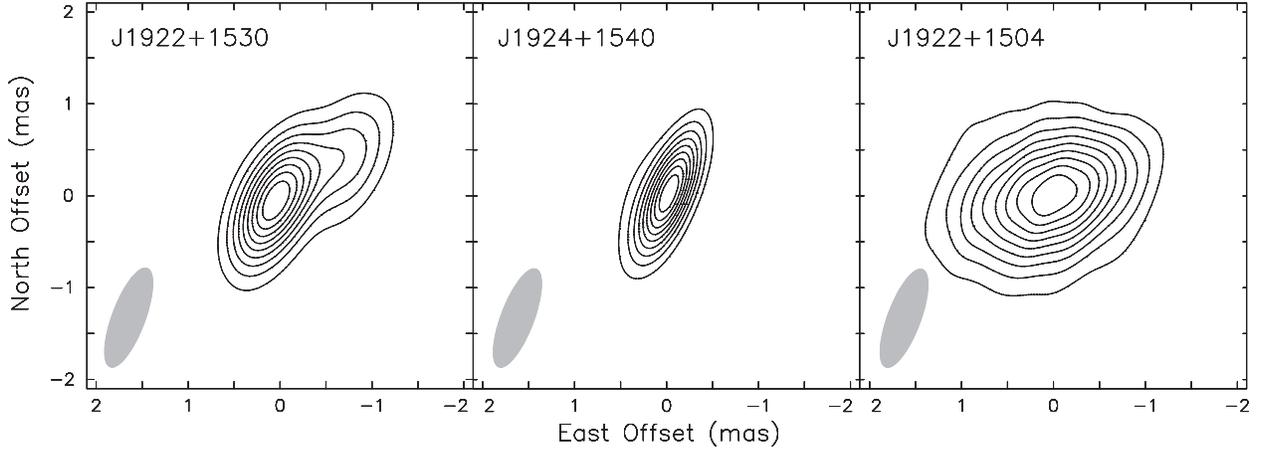} 
\caption{Images of background continuum sources.
Source names are shown in the upper left corner and restoring
 beams are indicated by gray ellipse in the lower left corner of
 each panel.  
The more heavily resolved source, J1922+1504, was not used
 for parallax measurement of \wsrc.
The contours are at 10\%, 20\%, 30\%, ..., 90\% of the peak
 intensities, which are 0.108 \jyb\ for J1922+1530, 0.633
 \jyb\ for J1924+1540, and 0.025 \jyb\ for J1922+1504.}
\label{fig:kntr}
\end{figure}

Since relative positions for two sources (\ie
 maser and quasar sources) at low antenna elevations are very
 sensitive to atmospheric delay errors \citep[\eg][]{Honma:08}, 
 we adopted an elevation
 cutoff of 30\deg\ for parallax measurement, below which data were
 discarded for each antenna.
At each epoch, we imaged the background continuum sources after
 calibration (Figure \ref{fig:kntr}) and then fitted elliptical
 Gaussian brightness distributions to the sources using the task
 JMFIT.
These positions were subtracted from the positions of the
maser spot at $\Vlsr=62.9$
 \kms\ in \ssrc\ (feature 32 in Table \ref{table:intmot}).
Since this maser spot was the phase reference at the first through 
third epochs, it had essentially zero position.  However, 
at the fourth epoch, this spot was not at the map center
since the phase reference was a different maser spot.  Since
both the maser spot at $\Vlsr=62.9$ \kms\ and the continuum sources 
are equally affected by the change in reference feature, differencing
their measured positions removes the effects of a different phase
reference spot in the last epoch.

The change in position of the maser spot at $\Vlsr=62.9$ \kms\
 relative to background continuum sources
 was then modeled by the parallax sinusoid in
 east-west ($\Delta \alpha \cos \delta$) and north-south
 ($\delta$) directions and a linear proper motion.  
This requires five parameters, \ie a single parallax parameter 
and two parameters for the proper motion in each coordinate.
Independent ``error floors'' were added to positions in each
 coordinate to allow for unknown systematic errors and then adjusted
 iteratively to make the reduced $\chi^2 \approx1.0$ in each
 coordinate \citep{Reid:09I}.

Figure \ref{fig:kntr} shows contour maps of the background
 sources.
Since one of the three background continuum sources, J1922+1504,
 displayed heavily resolved structure 
 ($\approx0.3$ mas), we did not use this source for parallax
 measurement.
For the two quasars, J1922+1530 and J1924+1540, after individual
 parallax fitting, we
 also attempted a ``combined'' fitting with the data from both
 quasars, solving for seven parameters, \ie a common parallax and
 proper motion of the maser source against
 both background quasars but different position
 offsets for each quasar.

\begin{figure}
\epsscale{1.} 
\plotone{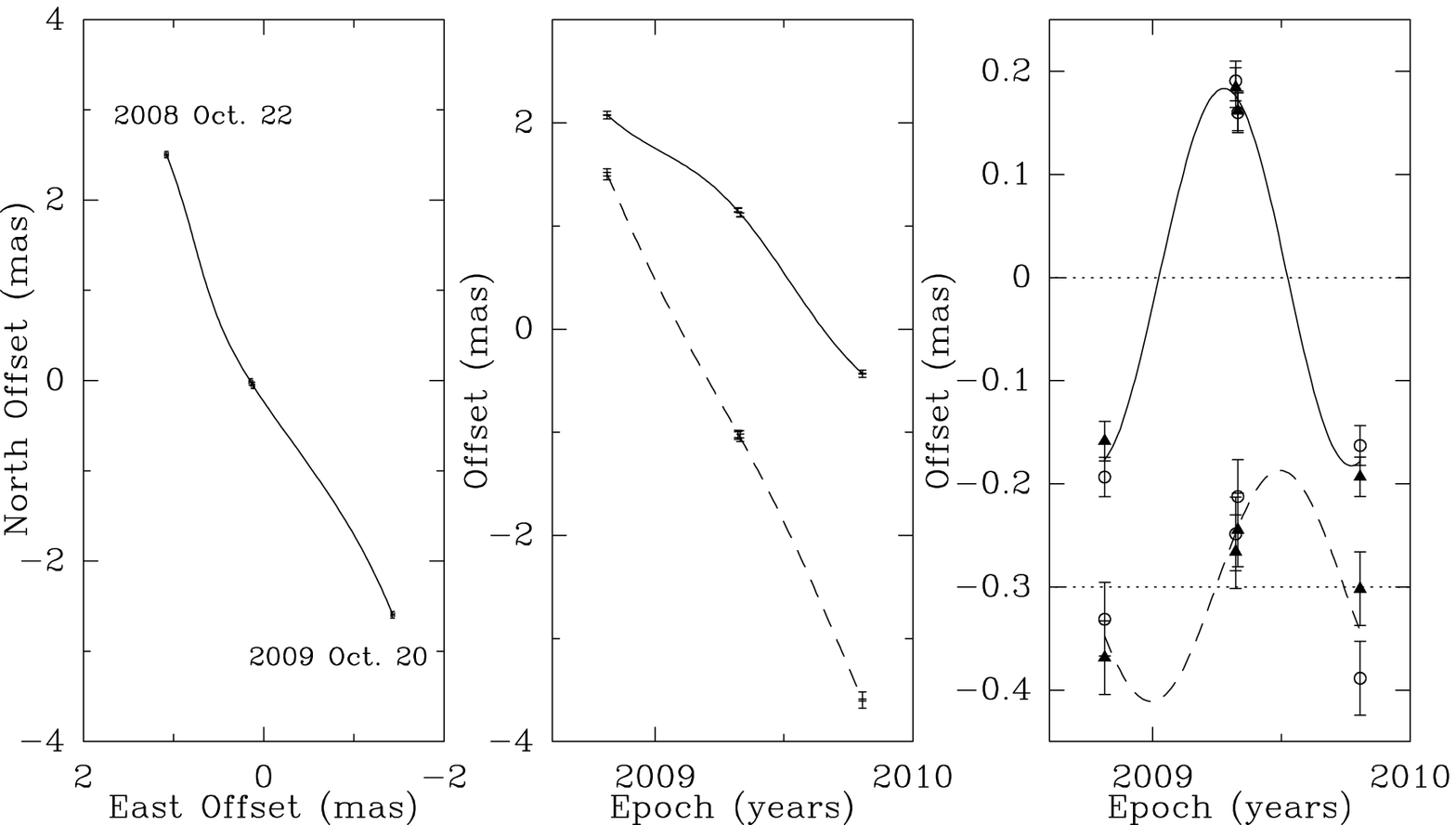} 
\caption{Parallax and proper motion data (points with error bars)
 and fits (solid or dashed curves) for \wsrc.
Plotted are position measurements of an \hho maser spot at
 $\Vlsr=62.9$ \kms\ relative to background quasars J1922+1530 and
 J1924+1540 with first and fourth epochs labeled.
Left panel: sky-projected motion of the maser spot relative to the
 two quasars.
Note that the second and third epochs were spaced only by 3 days
 and thus indistinguishable to eye in the plot.
Middle panel: east and north position offsets and the best-fit
 parallax and proper motion of the maser spot vs.\ time.  
Solid and dashed curves indicate fits for eastward and northward
 positions, respectively. 
Constant positional offsets from data are added in each coordinate
 for clarity.
Right panel: same as the middle panel, except the best-fit proper
 motion has been removed to show the effects of the parallax only.
Filled triangles show maser positions measured relative to
 J1922+1530 and open circles to J1924+1540. 
The northward data have been offset from the eastward data for
 clarity.}
\label{fig:parallax}
\end{figure}

Figure \ref{fig:parallax} shows position measurements and
 parallax fits of the masers relative to the two background sources,
 J1922+1530 and J1924+1540. 
The results of parallax fitting are listed in Table
 \ref{table:parallax}.
We obtained the parallax of \wsrc\ to be $\pi = 0.183\pm0.006$~mas
 using J1922+1530 and $\pi = 0.187\pm0.009$~mas using J1924+1540.
From the combined fit, we obtained $\pi = 0.185\pm0.007$~mas.
The uncertainty of each parallax fit was obtained from the formal
 fitting uncertainty.
The adjusted error floors which yielded $\chi^2 \approx1.0$
 were $\sigma_x = 0.014$~mas in the east-west direction and
 $\sigma_y = 0.023$~mas in the north-south direction using
 J1922+1530, and $\sigma_x = 0.020$~mas
 and $\sigma_y = 0.037$~mas using J1924+1540.
 For the combined fit, $\sigma_x = 0.020$~mas and
 $\sigma_y = 0.038$~mas.

We also attempted parallax fitting using other maser spots in both
 \msrc\ and South, which yielded essentially the same parallax
 as the maser spot at $\Vlsr=62.9$ \kms\ in \ssrc.
This indicates that structural changes in the maser spots
 are unlikely to be a significant error source for
 our parallax measurement of \wsrc. 
The dominant error source of the position measurements is likely to
 be residuals in the atmospheric delay modeling, but structural
 changes in background quasars may also add some ``noise.''
Random errors due to thermal noise are negligibly small,
 considering the high signal-to-noise ratio in both the maser and
 quasar maps.

The error-weighted mean of the two measurements using each quasar
 is $0.184\pm0.005$~mas.
The deviation of each quasar measurement from the mean,
 $\lesssim 0.003$~mas gives an estimate of the error intrinsic to
 each quasar, including potential quasar structural changes and 
 atmospheric delay error differences between the maser and quasar 
 positions.
In the right panel of Figure \ref{fig:parallax}, the positions
 in the east-west direction at the second and third
 epochs spaced by 3 days differ by $\approx 0.02$ mas
 from each other, while the measurements on the same day using
 different quasars differ by a much smaller amount
 $\approx 0.003$ mas.
The position difference between 3 days is likely due to the
 atmospheric delay errors, since structural change of the quasars
 over 3 days is unlikely.
Therefore, the dominant source of the position errors
 appears to be the residuals in the atmospheric delay modeling.

Since the two quasars lie relatively near each other on the sky
 (see Table \ref{table:pos} for angular separations and position
 angles relative to \wsrc), the atmospheric delay errors may be
 partially correlated for the two quasars.
The uncertainty of the combined-fit parallax, estimated from 
 the formal fitting uncertainty, assumes that the errors of the
 measurements using two quasars are uncorrelated.
If they are 100\% correlated, the estimated error needs to be
 multiplied by $\sqrt{(16-7)/(8-5)}=\sqrt{3}$
 (\ie 16 data points and 7 parameters yielding 9 degrees of
 freedom for uncorrelated data, and 8 data points and
 5 parameters for a single quasar case).
As an estimate of partially correlated errors,
 we average the factors of 1 (for
 no correlation) and $\sqrt{3}$ (for a 100\% correlation),
 and multiply the formal error by
 $(1+\sqrt{3})/2\approx 1.37$, which yields an uncertainty
 of $\sigma_{\pi} = 0.010$~mas, as the best estimate
 for partially correlated data.
We thus adopt the combined-fit parallax of
 $\pi = 0.185 \pm 0.010$~mas as the best estimate, which
 corresponds to a source distance of
 5.41$^{+0.31}_{\,-0.28}$~kpc.
 
Our parallax distance is in good agreement with
 5.1$^{+2.9}_{\,-1.4}$~kpc for \isrc\ (or North)
 by \citet{Xu:09} from 12 GHz \meth
 maser parallax, but has much higher accuracy.
The parallax by \citet{Xu:09} has much larger uncertainty
 because the 12 GHz masers in \isrc\ had large spot sizes
 and were resolved out at longer baselines of the VLBA,
 which resulted in lack of data at the highest resolution.
Also, the low signal-to-noise ratio due to the low 12 GHz maser
 flux density may have contributed to the error of the
 measurement.
Previous to the parallax measurements, the distance to
 W51 was mostly based on statistical parallaxes based on
 22 GHz \hho maser proper motions, \eg 7$\pm 1.5$~kpc by
 \citet{Genzel:81} for \msrc\ and 8.3$\pm 2.5$~kpc by
 \citet{Schneps:81} and 6.1$\pm 1.3$~kpc by \citet{Imai:02}
 for \nsrc.
\wsrc\ and \isrc/North are separated by
 $\approx 1{\rlap{.}}{'}0$ east and $0{\rlap{.}}{'}6$ south.
Both are thought to be associated with the \hii region W51~A.
Our distance is consistent with the upper limit of 5.8~kpc by
 \citet{Barbosa:08} from spectroscopic study of \isrc, but is in
 significant disagreement with the smaller distance of
 2.0$\pm 0.3$ kpc recently reported by \citet{Figueredo:08} from
 spectroscopic parallaxes of four O-type stars near the giant \hii
 region W51~A.
\citet{Figueredo:08} discuss an extra uncertainty in the derived
 distance that comes from the interstellar extinction law, however
 a factor of 1.28 difference in distance they cite for adopting
 different methods to estimate $A_K$ will not account for 
 the distance discrepancy.
As discussed by \citet{Clark:09}, the underestimated spectroscopic
 distance by \citet{Figueredo:08} may come from
 an underestimate of the luminosity class 
 of the stars (O4 to O7.5) used to derive the distance. 
Our distance of 5.4~kpc suggests that the luminosity of the P~Cyg
 supergiant type star [OMN2000] LS1, whose nature \citet{Clark:09}
 discuss for three different assumptions on its distance, is
 indeed very high and of order $\sim 5\times 10^5~L_\odot$.


\subsection{Maser Map and Internal Motions} \label{sect:intmot}

We detected 37 \hho maser features in \wsrc\ that we could identify
 at all four epochs, and we list their relative proper motions
 in Table \ref{table:intmot}.
We classified features into four separate groups according to their
 likely exciting sources: hyper-compact (HC) or ultra-compact (UC)
 \hii regions W51e2-NW,
 W51e2-E \citep{Shi:10, Shi:10II} and W51e8 \citep{Zhang:97}, and a region
 we call ``u'' without detected millimeter or
 centimeter continuum emission (see Figure \ref{fig:intmot}).
Note that a proper motion of 2 \masy\ corresponds to 51 \kms\ at
 the distance of 5.4 kpc.

Figure \ref{fig:intmot} gives a map of positions and internal
 motions of the detected maser features.
The origin of the map is the position of the reference
 feature 32.
The plotted internal motions are relative to the average velocity
 of the four regions,
 $\langle \Delta v_x \rangle = -3$ \kms\
 and $\langle \Delta v_y \rangle = 1$ \kms\ 
 relative to the maser feature 32 (see Section \ref{sect:mean}).
The positions of the HC and UC \hii regions are also
 plotted by crosses whose sizes approximate
 the beam sizes from previous observations:
 W51e1, e3 and e4 by \citet{Gaume:93}; W51e8 by \citet{Zhang:97}
 and W51e2-W, e2-E, e2-NW and e2-N by \citet{Shi:10}.
Molecular-line studies show evidence for gravitational collapse of
 W51e2 (\msrc) and W51e8 (\ssrc) cores \citep{Ho:96, Zhang:98,
 Tang:09}.

For the W51e2 region, we can identify two outflows:
 a high-velocity outflow and a low-velocity outflow
 that arise separately from the e2-E and e2-NW cores, respectively.
We find that the northwest to southeast direction of the bipolar \hho maser
 outflow from W51e2-E is in good agreement with the
 molecular outflow direction seen in the CO(2--1) line \citep{Keto:08II}
 and the CO(3--2) and HCN(4--3) lines \citep{Shi:10, Shi:10II},
 but with higher expansion velocities for \hho masers.
We classified maser features near W51e2 position into e2-E and
 e2-NW groups by their radial velocities: features with radial
 velocities within the rage of $0 < \Vlsr < 120$~\kms\ were
 included in e2-NW and high-velocity features outside of the range
 in e2-E.
For comparison, the molecular radial velocity of W51e2 was
 found to be 55~\kms\
 by \citet{Zhang:98} from CH$_3$CN (8-7) line observations.


\section{Discussion} 

\subsection{Distance to the Galactic Center}
\label{sect:R0}

Since the location of W51 is very close to the tangent point
 of the Sagittarius spiral arm of the Galaxy,
 the distance to W51 directly yields a good estimate of
 the distance \Ro to the Galactic center by simple geometry.  

\begin{figure}
\epsscale{0.7} 
\plotone{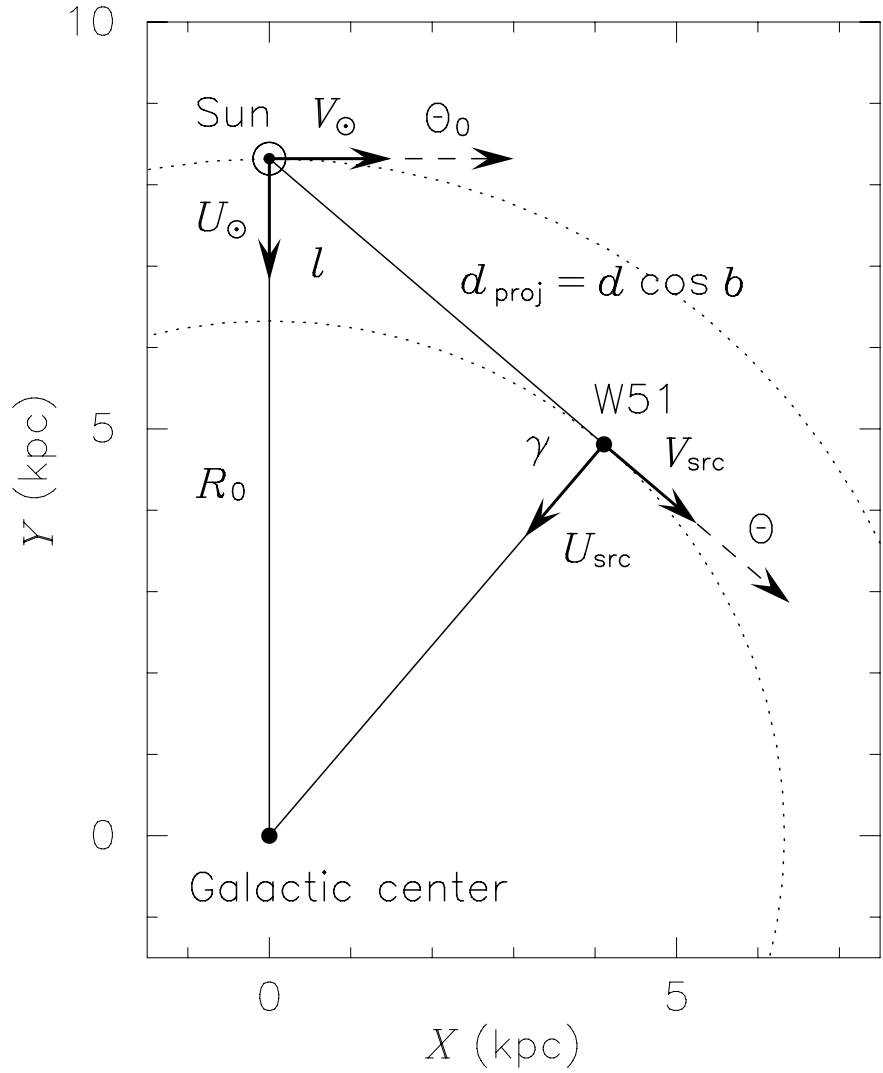} 
\caption{Schematic depiction of source locations
 and Galactic parameters when the source is at
 the ``tangent point'' ($\gamma=90^\circ$) in the Galaxy.}
\label{fig:R0}
\end{figure}

Figure \ref{fig:R0} gives a schematic depiction of the source
 locations.
At the galactic coordinates of 
 ($l$, $b$)$=$($49\d 49$, $-0\d 39$),
 W51 lies very close to the tangent point 
 (\ie the angle $\gamma$ between the Sun and
 the Galactic center as viewed from W51 is very close to $90^\circ$; 
 see Figure \ref{fig:R0}), since the radial velocity of \wsrc,
 $\Vlsr \approx 58$~\kms\ (Section \ref{sect:mean}),
 is very close to the maximum radial velocity
 (which occurs at the tangent point) given by
 $v_{\rm max} = \Theta_0 (1-\sin l)$,
 assuming a flat rotation curve of the Galaxy
 (\ie $\mbox{d}\Theta/\mbox{d}R = 0$).
For example, if we adopt 
$R_0 = 8.4\pm 0.4$~kpc \citep{Ghez:08, Gillessen:09}
 and $\Theta_0/R_0 = 29.45 \pm 0.15$ \kms~kpc$^{-1}$
 \citep{Reid:04}
 to obtain $\Theta_0$,
 then the maximum velocity for $l=49\d 49$ is
 $v_{\rm max} = 59\pm 3$~\kms,
 which is in good agreement with 
 the observed radial velocity of \wsrc\
 ($\langle \Vlsr \rangle = 58\pm 4$~\kms;
 see Section \ref{sect:mean}).
Note that correction for a likely counter-rotating peculiar
 (non-circular) motion of a massive star forming region,
 as reported by \citet{Reid:09VI}, would only make the observed
 radial velocity due to the Galactic rotation larger than the
 expected maximum velocity, which 
 would then lend further support to the assumption that
 W51 lies at the tangent point.   

From simple geometry, $R_0 = d \cos b / \cos l$, assuming only
 that W51 lies at the tangent point (\ie $\gamma = 90^{\circ}$),
 our distance estimate of $d = 5.41^{+0.31}_{\,-0.28}$~kpc to W51
 directly gives a trigonometric distance of
 $R_0 = 8.32^{+0.48}_{\,-0.43} \approx 8.32 \pm 0.46$~kpc
 to the Galactic center.

We now relax the assumption that W51 is exactly at the
 tangent point and estimate the systematic effect this has
 on the uncertainty of $R_0$.
For example, if we adopt Galactic and source parameters
 as discussed in Section \ref{sect:galaxy} (\ie
 $\To = 240 \pm 20$~\kms; $U_{\odot} = 10.00 \pm 0.36$~\kms,
 $V_{\odot} = 8.0 \pm 3.0$~\kms, $W_{\odot} = 7.17 \pm 0.38$~\kms; 
 the source proper motion $\mu_x = -2.64 \pm 0.16$~\masy,
 $\mu_y = -5.11 \pm 0.16$~\masy; the source radial velocity 
 $\Vlsr = 58 \pm 4$~\kms\ and the radial peculiar motion
 $v_{\rm rad} \approx V_{\rm src} = -4 \pm 5$~\kms) and solve for $\Ro$
 using the equation by \citet{Reid:88II} 
 with no assumption on the tangent point,
 we obtain $\Ro = 8.1 \pm 1.1$~kpc,
 which includes both statistical and systematic uncertainties.
Therefore our best estimate of \Ro from the parallax of W51 is
 $\Ro=8.3\pm 0.46$ (statistical) $\pm 1.0$ (systematic) kpc.
Combining the statistical and systematic uncertainties
 in quadrature, we find $\Ro = 8.3 \pm 1.1$~kpc.
This estimate is consistent with other measurements of $R_0$:
 \eg $R_0 = 8.0 \pm 0.5$~kpc by \citet{Reid:93}
 from an ensemble of classical techniques,
 and recent direct measurements of 
 $R_0 = 8.4 \pm 0.4$~kpc by \citet{Ghez:08} and
 $R_0 = 8.33 \pm 0.35$~kpc by \citet{Gillessen:09}
 from stellar orbits around \SgrA,
 $R_0 = 7.9^{+0.8}_{\,-0.7}$~kpc by \citet{Reid:09R0}
 from the trigonometric parallax of Sgr B2,
and $R_0 = 8.4 \pm 0.6$~kpc by \citet{Reid:09VI}
 from trigonometric parallaxes and proper motions of 16 massive
 star-forming regions across the Galaxy.
We thus determined the distance to the Galactic center
 with good accuracy from a precise parallax measurement of \wsrc.

\subsection{Origins of Internal Maser Motions}\label{sect:mean}

\begin{figure}
\epsscale{0.9} 
\plotone{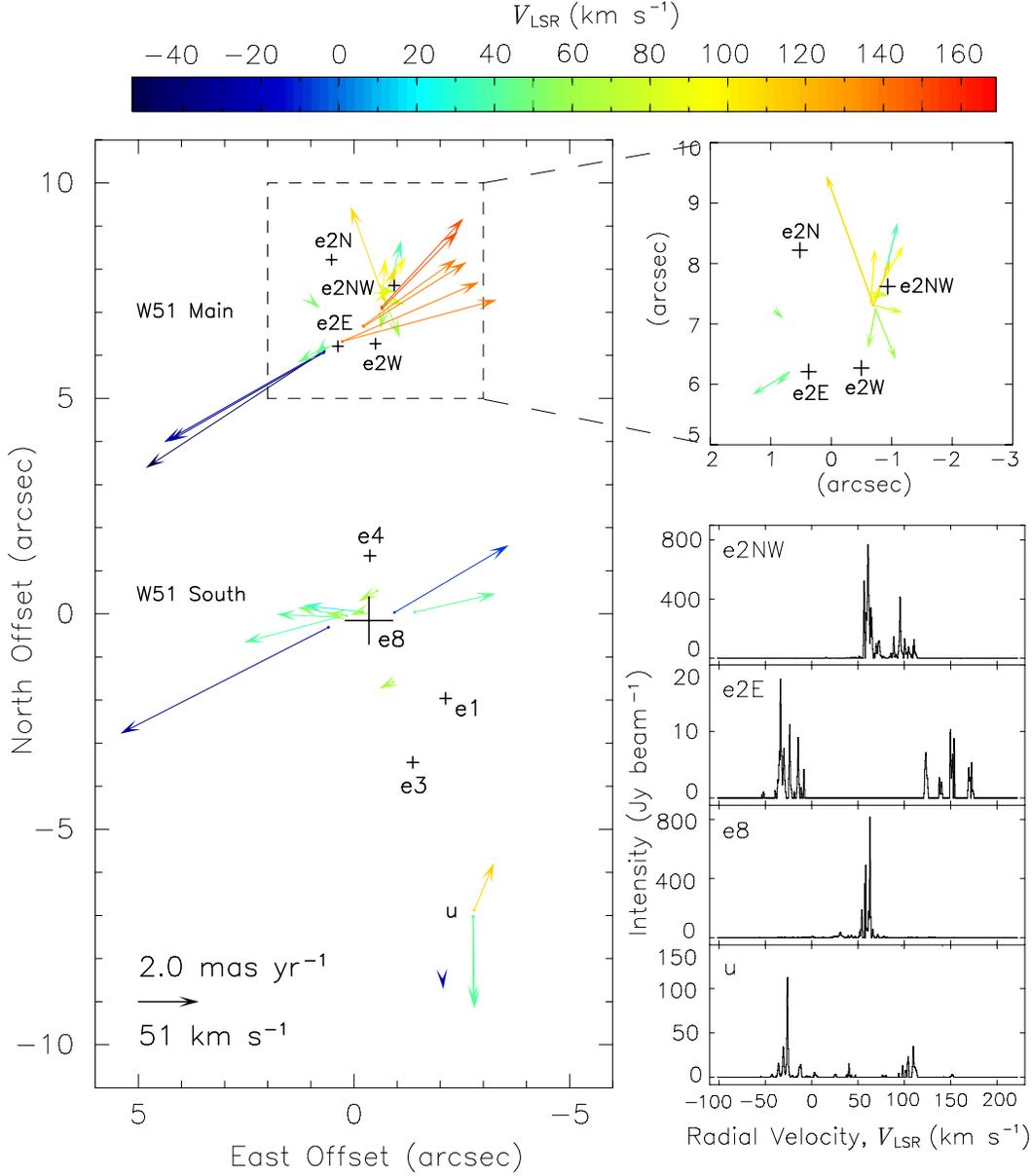} 
\caption{
Left panel: map of the positions and internal motions of \hho maser
 features in \wsrc, as seen from the mean motion of all regions
 (see Section \ref{sect:mean}).  
Absolute coordinates of the map origin 
 are R.A.\ (J2000)$=19^{\rm h} 23^{\rm m}
 43{\rlap{.}}^{\rm s} 93427$ and
 decl.\ (J2000)$=14^{\circ} 30{'} 28{\rlap{.}}{''} 3498$
 (\ie position of reference feature 32 at the
 mid-observation year of 2009.318).
The HC and UC \hii regions are
 marked by crosses with positions and beam sizes from previous
 observations (see the text for references).
A sample motion of 2.0 \masy\ ($\approx$51 \kms\ at a distance of
 5.4 kpc) is shown at the bottom left corner of the map.
Right panel (top): blowup for W51e2-NW region with lower-velocity
 maser spots.
Right panel (bottom): \hho maser spectra of the four regions with
 intensities summed over all spots detected at the first epoch.
}\label{fig:intmot}
\end{figure}

The internal proper motions of the \hho masers in \wsrc\
 (Figure \ref{fig:intmot}) suggest four bipolar outflows,
 presumably from four young, massive stars: e2-NW, e2-E,
 e8, and u.  
We modeled the first three as uniformly expanding flows and
 fitted the data to obtain seven global parameters: the
 expansion speed and the three spatial and
 velocity components of the center of expansion.
(There were not enough measured motions for the ``u'' group to
 reliably model.) 
For each maser feature, we fitted the measured three
 components of velocity using their positions on the sky
 as independent variables.  
Since the spatial offset along the line of sight from the
 center of expansion for each maser feature is unknown, we
 needed to solve for an additional offset parameter for
 each feature.

We adopted a Bayesian fitting procedure using a Markov chain
 Monte Carlo method to explore parameter space. 
\hho masers are known to have an internal dispersion of
 $\sim15$ \kms\ for each velocity component \citep{Reid:88}
 and we added this in quadrature with the (usually much
 smaller) measurement errors.  
Anticipating that some maser features may be ``outliers''
 (\ie not well modeled by simple uniform expansion),
 we adopted an ``error-tolerant'' probability density
 function for the data errors
 with the form ${\rm prob}(\sigma|\sigma_0) = \sigma_0/\sigma^2$,
 where $\sigma_0$ is the adopted (minimum) data uncertainty
 (see \citealt{Sivia:07}).
For priors, we set the expansion speed ($V_{\rm exp}$) at
 $75\pm200$~\kms; the $x$ and $y$ proper motions
 of the center ($\Delta v_x$ and $\Delta v_y$) at $0\pm10$~\kms\
 relative to the reference feature \#32;
 the LSR velocity of the center (\Vlsr) was taken as $57\pm10$
~\kms, a value between the average of the thermal molecular
 lines in e2 and e8 ($\Vlsr =$ 55 and 59 \kms\ for e2
 and e8, respectively, by \citealt{Zhang:98} from the
 CH$_3$CN (8--7) line);
 the coordinates of the center of expansion ($x$ and $y$) were
 estimated visually from Figure \ref{fig:intmot} and assigned
 an uncertainty of $\pm1$ arcsec
 (\ie $(x,~y)=(0\pm 1,~6.5 \pm 1)$ arcsec for e2-NW,
  $(0.3\pm 1,~6.2 \pm 1)$ arcsec for e2-E, and
  $(-0.5\pm 1, 0\pm 1)$ arcsec for e8); and we assumed a source
 distance given by the parallax ($\pi$) of $0.185\pm 0.010$~mas
 determined in Section \ref{sect:parallax}.  

The results of the fitting for each center of expansion 
 are given in Table \ref{table:average}.
We adopted the peak value of the histogram of each parameter
 as the best estimate of the parameter,
 and estimated its uncertainty
 from the halfwidth of the histogram where the number of
 occurrences becomes $\exp(-1/2)$ of the peak,
 \ie 1$\sigma$ for a Gaussian distribution.
The error-weighted means of the fitted velocities are
 $\langle \Delta v_x \rangle = -3 \pm 4$~\kms\ and
 $\langle \Delta v_y \rangle =  1 \pm 4$~\kms\
 relative to feature 32, and 
 $\langle \Vlsr \rangle = 58 \pm 4$~\kms.
These values are in good agreement
 with the unweighted mean of
 all 37 maser features in Table \ref{table:intmot}:
 $\langle \Delta v_x \rangle = 0$~\kms\ and
 $\langle \Delta v_y \rangle = 2$~\kms\ (at a
 distance of 5.41~kpc) and $\langle \Vlsr
 \rangle = 62$~\kms.
We adopt the result from the Bayesian method
 as the best estimate for the systemic velocity of \wsrc.
By adding
 $\langle \Delta v_x \rangle = -3 \pm 4$~\kms\ and
 $\langle \Delta v_y \rangle =  1 \pm 4$~\kms\
 (at a distance of 5.41~kpc) to the absolute
 proper motion of the feature 32,
 $(\mu_x,~\mu_y)=(-2.53 \pm 0.02,~-5.15 \pm 0.04)$ \masy\
 (see Section \ref{sect:parallax} and
 Table \ref{table:parallax}), we obtain the systemic
 absolute proper motion of \wsrc\ to be 
 $\langle \mu_x \rangle= -2.64 \pm 0.16$~\masy\ and
 $\langle \mu_y \rangle= -5.11 \pm 0.16$~\masy,
 which will be used in Section \ref{sect:galaxy} for fitting 
 Galactic parameters. 
Figure \ref{fig:intmot} shows the maser motions relative
 to this systemic velocity.

For comparison, \citet{Xu:09} report proper motions of 
 $(\mu_x, \mu_y)=(-2.49 \pm 0.07$, $-5.45 \pm 0.14)$
 \masy\ and 
 $(-2.48 \pm 0.08,$ $-5.56 \pm 0.08)$ \masy\
 for two \meth maser spots in \isrc, whose sky position is 
 ${1\rlap{.}}{'}0$ west and $0{\rlap{.}}{'}6$ north
 of \wsrc\ (see, \eg \citealt{Figueredo:08} for a map
 of the giant \hii region W51~A).
At the source distance of 5.4~kpc, 
 the two regions are separated by a sky-projected
 distance of $\approx 1.8$~pc.
The eastward proper motions $\mu_x$ of the two regions agree
 well, and the northward proper motions $\mu_y$ differ by 
 $\approx 0.4$~\masy $\approx 10$~\kms. 
The difference in $v_y$ of 10~\kms\ is a reasonable value, 
 considering the virial motion of $\approx 7$~\kms\ in each
 coordinate expected for each massive star that is 
 exciting the maser emission \citep{Reid:09VI}
 and some contribution from the internal motion
 of the \meth maser spots $\approx 3$~\kms\
 \citep{Moscadelli:02}.

Each \hho maser outflow in \wsrc\ shows
 a relatively small range of internal 3D speeds.
This is contrary to previous observations of
 other Galactic star-forming regions,
 \eg Orion-KL \citep{Genzel:77, Genzel:81a}
 and W49N \citep{Gwinn:92}, which suggested that
 high- and low-velocity outflows might originate from
 the same young stellar object.
In W49, a strong acceleration region at a radius of
 0.1~pc is required if all masers are associated
 with a single young stellar object.
However, these conclusions were based on observations with 
 less precise maser motions and, most importantly,
 much lower resolution dust emission maps.
Our results for W51 suggest that multiple speed outflows
 may come from separate young stellar objects that can be
 very closely spaced on the sky.

\subsection{Galactic 3D Motion of W51}
\label{sect:galaxy}

Using our parallax distance and
 the fitted full-space systemic velocity of \wsrc\
 obtained in Section \ref{sect:mean}, we now estimate the
 3D motion of \wsrc\ in the Galaxy.
As in Section \ref{sect:mean}, we adopted a Bayesian fitting procedure
 using a Markov chain Monte Carlo method.
We set priors for nine parameters: 
 the source distance of \wsrc\ given by the parallax ($\pi$)
 determined in Section \ref{sect:parallax};
 distance to the Galactic center ($R_0$) and
 the angular rotation speed ($\Theta_0 / R_0$) of the Galaxy 
 at the LSR;
 the Solar motion ($U_{\odot}$, $V_{\odot}$, $W_{\odot}$)
 relative to the LSR; and source peculiar motion
 (relative to a circular orbit) of \wsrc\
 ($U_{\rm src}$, $V_{\rm src}$, $W_{\rm src}$),
 where $U$, $V$, and $W$ denote velocity components in
 the directions (at the source location)
 toward the Galactic center,
 toward Galactic rotation
 and toward the north Galactic pole (NGP), respectively.

The prior $R_0 = 8.25 \pm 0.4$~kpc was chosen as
 an average between $8.0$~kpc by \citet{Reid:93}
 and the recent direct measurements of \Ro
 \citep[\eg][]{Ghez:08, Gillessen:09}.
The prior value of $\To/\Ro = 29.45 \pm 0.15$~\kms~kpc$^{-1}$ 
 (for $V_{\odot} = 5.25$~\kms\
 and corrected for the varied $V_{\odot}$ values)
 was adopted from \citet{Reid:04} from the proper motion of \SgrA.  
The priors of $U_{\odot} = 10.00 \pm 0.36$~\kms\ and
 $W_{\odot} = 7.17 \pm 0.38$~\kms\ were adopted from
 {\it Hipparcos} values by \citet{Dehnen:98}, while
 $V_{\odot} = 8.0 \pm 3.0$~\kms\
 was chosen to allow for both 5.25~\kms\ by \citet{Dehnen:98}
 and a larger value of $\sim 11$~\kms\
 recently revised by \citet{Binney:10}. 
We set the priors of
 $U_{\rm src}$, $V_{\rm src}$ and $W_{\rm src}$
 to be rather open, $0 \pm 20$~\kms, in particular to allow for 
 a wide possible range of $V_{\rm src}$ due to the uncertainty
 of $V_{\odot}$ \citep{Reid:09VI, Binney:10, McMillan:10,
 Schoenrich:10}.
We assumed a flat rotation curve of the Galaxy
 (\ie $\mbox{d}\Theta/\mbox{d}R = 0$).
The nine parameters were then adjusted
 to maximize the probability of the model given the above priors
 and the data of the source proper motion and LSR velocity,
 $\langle \mu_x \rangle= -2.64 \pm 0.16$~\masy,
 $\langle \mu_y \rangle= -5.11 \pm 0.16$~\masy\ and 
 $\langle \Vlsr \rangle= 58\pm 4$~\kms.
Note that the proper motions ($\mu_x,~\mu_y$) are
 in the heliocentric frame, while the given radial velocity
 $\Vlsr$ is the LSR velocity,
 converted from the heliocentric
 radial velocity $v_{\rm helio}$
 (obtained from the Doppler shift) using 
 the standard solar motion by definition 
 relative to the LSR (see \citealt{Reid:09VI}).

Since the obtained probability distribution functions
 for some parameters did not follow a Gaussian function, 
 we adopted, as a conservative approach,
 the mean value of each parameter
 as the best estimate for the parameter
 and the standard deviations from the mean
 as the uncertainty, 
  based on the last 90\% of 500,000 trials.
Most of the Galactic parameters returned
 the uncertainties only slightly smaller than the priors,
 indicating that these parameters were
 mainly constrained by the fairly strong priors.

The fitted result yields
 ($U_{\rm src}$, $V_{\rm src}$, $W_{\rm src}$)$=$
 ($-7\pm 5$, $-4 \pm 5$, $4\pm 4$)~\kms\ for 
 the peculiar (non-circular) motion of \wsrc.
For comparison, the parallax and full-space motion
 reported by \citet{Xu:06} for
 W3(OH) in the Perseus spiral arm would yield
 ($U_{\rm src}$, $V_{\rm src}$, $W_{\rm src}$)$=$
 ($19\pm 3$, $-15 \pm 4$, $1\pm 2$)~\kms\ 
 under the same priors for Galactic parameters.
Therefore, the full-space velocity of \wsrc\ suggests that
 W51 is in a nearly circular orbit about the Galactic center,
 with no large peculiar motion.

\acknowledgements
The authors thank an anonymous referee
 for many valuable comments.
MS acknowledges financial support from JSPS Research Fellowships
 for Young Scientists.
This work was supported by Grant-in-Aid for JSPS Fellows and
 conducted as part of her visiting research at
 Harvard-Smithsonian Center for Astrophysics through
 Smithsonian Astrophysical Observatory Predoctoral Program and
 JSPS Excellent Young Researchers Overseas Visit Program.

{\it Facilities:} \facility{VLBA}


\begin{deluxetable}{lllllll}
\tablecolumns{7} \tablewidth{0pc} 
\tablecaption{VLBA Observations of \wsrc}
\tablehead {
  \colhead{Epoch} & \colhead{Date} & \colhead{Day of Year} & 
  \colhead{Time Range} & \colhead{Restoring Beam} &
  \multicolumn{2}{c}{Antennas Unavailable}
\\
\cline{6-7} \\
  \colhead{}      & \colhead{}     & \colhead{} & 
  \colhead{(UT)}       & \colhead{(mas, mas, deg)} &
  \colhead{No Data} & {Flagged}
            }
\startdata
 1 & 2008 Oct 22 &2008/296 & 19:36-04:31 & 1.2$\times$0.4 at --20 & SC & HN \\
 2 & 2009 Apr 27 &2009/117 & 07:20-16:16 & 0.8$\times$0.3 at ~--7 & ...& NL \\
 3 & 2009 Apr 30 &2009/120 & 07:09-16:04 & 1.1$\times$0.3 at --18 & KP & NL, SC\\
 4 & 2009 Oct 20 &2009/293 & 19:44-04:40 & 1.1$\times$0.3 at --19 & SC & ... \\
\enddata
\tablecomments {
Restoring beam sizes (FWHM) are indicated in Column 5 by the major
 and minor axes in mas and position angles (P.A.) east of north in
 degrees.
Antenna codes are HN: Hancock, KP: Kitt Peak, NL: North Liberty,
 and SC: Saint Croix.} 
\label{table:obs}
\end{deluxetable}

\begin{deluxetable}{lllcrlc}
\tablecolumns{7} \tablewidth{0pc} \tablecaption{Source Positions
 and Brightnesses}
\tablehead {
  \colhead{Source} & \colhead{R.A. (J2000)} &
  \colhead{Dec. (J2000)} &  \colhead{$\theta_{\rm sep}$} &
  \colhead{P.A.} &  \colhead{$T_b$} & \colhead{\Vlsr} 
\\
  \colhead{}       & \colhead{(h~~m~~s)} &
  \colhead{(d~~$'$~~$''$)} &  \colhead{(deg)}  & \colhead{(deg)} &
  \colhead{(Jy/b)} & \colhead{(\kms)}
            }
\startdata
 \wsrc       & 19 23 43.87363  &+14 30 29.4536  & ...  & ...   & 746    & 58 \\
 J1922+1530  & 19 22 34.699314 &+15 30 10.03262 & 1.0  & $-16$ & 0.108  & ...  \\
 J1924+1540  & 19 24 39.455870 &+15 40 43.94172 & 1.2  & $11$  & 0.633  & ...  \\
 J1922+1504  & 19 22 33.27291  &+15 04 47.5382  & 0.6  & $-26$ & 0.025  & ...  \\
\enddata
\tablecomments {  
Coordinates are those used in the VLBA correlator.  
Absolute positions of J1922+1530 and J1924+1540 are based on the
 VLBA Calibrator Survey 1 
\citep[VCS1; ][]{Beasley:02} and known in each coordinate to about
 $\pm0.3$~mas (J1922+1530) and $\pm0.8$ mas (J1924+1540).
The position of J1922+1504 is accurate to about $\pm20$~mas.
Angular offsets ($\theta_{\rm sep}$) and position angles (P.A.)
 east of north relative to the \hho maser source in \wsrc\ are
 indicated in Columns 4 and 5. 
Peak brightnesses ($T_b$) for the maser and background continuum
 sources are from the first epoch.
  }
\label{table:pos} 
\end{deluxetable}

\begin{deluxetable}{rrrrrrr}
\tablecolumns{7} \tablewidth{0pc} 
\tablecaption{Maser Feature Motions in \wsrc}
\tablehead {
  \colhead{ID} & \colhead{\Vlsr} & \colhead{$T_b$} &
  \colhead{$x_0$} & \colhead{$y_0$} &
  \colhead{$\mu_x$}    & \colhead{$\mu_y$} 
\\
  \colhead{}           & \colhead{(\kms)} & \colhead{(Jy/b)}&
  \colhead{($''$)}     & \colhead{($''$)} &
  \colhead{(\masy)}    & \colhead{(\masy)}
            }
\startdata
 (e2-NW)\\
   1 &    39.3 &     1.8 &  --0.879 (0.008)  &    7.786 (0.025)  & --0.43 (0.03)  &    1.31 (0.11)  \\
   2 &    45.0 &     4.2 &    0.721 (0.008)  &    6.177 (0.008)  &   0.14 (0.02)  &  --0.23 (0.02)  \\
   3 &    46.6 &     9.3 &    0.700 (0.005)  &    6.196 (0.003)  &   0.75 (0.01)  &  --0.49 (0.01)  \\
   4 &    51.4 &     6.0 &    0.930 (0.023)  &    7.216 (0.001)  & --0.32 (0.06)  &  --0.14 (0.01)  \\
   5 &    60.7 &   644.5 &  --0.732 (0.010)  &    7.238 (0.005)  & --0.60 (0.02)  &  --1.15 (0.01)  \\
   6 &    61.9 &   336.9 &  --0.732 (0.006)  &    7.238 (0.024)  &   0.06 (0.02)  &  --0.87 (0.06)  \\
   7 &    72.5 &   114.9 &  --0.674 (0.003)  &    7.518 (0.003)  & --0.53 (0.01)  &  --0.04 (0.01)  \\
   8 &    77.5 &     2.5 &  --0.674 (0.044)  &    7.315 (0.023)  & --0.85 (0.11)  &  --0.14 (0.05)  \\
   9 &    83.6 &    13.6 &  --0.662 (0.020)  &    7.416 (0.021)  & --0.41 (0.09)  &    0.14 (0.10)  \\
  10 &    95.5 &   357.5 &  --0.655 (0.003)  &    7.367 (0.002)  & --0.89 (0.01)  &    1.37 (0.01)  \\
  11 &   100.1 &   601.3 &  --0.656 (0.006)  &    7.361 (0.003)  & --0.21 (0.01)  &    1.30 (0.01)  \\
  12 &   106.8 &    73.4 &  --0.702 (0.007)  &    7.306 (0.012)  & --0.51 (0.04)  &    1.11 (0.12)  \\
  13 &   109.5 &   103.2 &  --0.701 (0.061)  &    7.307 (0.087)  &   1.00 (0.13)  &    3.11 (0.18)  \\
 \hline  
 (e2-E)~~~~\\
  14 &  --51.7 &     3.6 &    0.800 (0.011)  &    6.033 (0.030)  &   5.64 (0.02)  &  --3.74 (0.07)  \\
  15 &  --33.9 &    18.1 &    0.706 (0.013)  &    6.059 (0.006)  &   5.01 (0.04)  &  --2.90 (0.02)  \\
  16 &  --23.6 &    34.2 &    0.687 (0.003)  &    6.095 (0.005)  &   5.20 (0.02)  &  --2.98 (0.01)  \\
  17 &   137.9 &     4.5 &  --0.210 (0.015)  &    6.683 (0.003)  & --3.22 (0.04)  &    2.26 (0.01)  \\
  18 &   138.5 &     2.7 &    0.269 (0.008)  &    6.323 (0.011)  & --5.24 (0.02)  &    1.42 (0.02)  \\
  19 &   139.6 &     1.0 &  --0.229 (0.008)  &    6.670 (0.004)  & --3.53 (0.03)  &    2.18 (0.01)  \\
  20 &   139.8 &    12.0 &    0.269 (0.016)  &    6.323 (0.013)  & --4.65 (0.04)  &    2.01 (0.03)  \\
  21 &   149.8 &     6.2 &  --0.646 (0.007)  &    7.125 (0.010)  & --2.83 (0.03)  &    2.98 (0.03)  \\
  22 &   150.9 &     3.5 &  --0.649 (0.017)  &    7.083 (0.017)  & --2.64 (0.05)  &    2.61 (0.05)  \\
 \hline
 (e8)~~~~~~\\
  23 &  --28.7 &    10.0 &    0.591 (0.001)  &  --0.311 (0.005)  &   6.78 (0.01)  &  --3.49 (0.02)  \\
  24 &   --7.4 &     5.5 &  --0.938 (0.002)  &    0.034 (0.010)  & --3.90 (0.01)  &    2.27 (0.02)  \\
  25 &    26.6 &     6.7 &  --0.198 (0.005)  &    0.049 (0.007)  &   1.84 (0.02)  &    0.25 (0.02)  \\
  26 &    41.3 &    81.7 &    0.365 (0.073)  &  --0.074 (0.075)  &   3.01 (0.15)  &  --0.80 (0.16)  \\
  27 &    43.0 &    37.3 &  --1.404 (0.003)  &    0.039 (0.005)  & --2.81 (0.01)  &    0.66 (0.02)  \\
  28 &    42.3 &    11.5 &    0.365 (0.027)  &  --0.074 (0.020)  &   1.93 (0.07)  &    0.13 (0.06)  \\
  29 &    52.1 &    27.8 &    0.172 (0.012)  &  --0.040 (0.023)  &   1.52 (0.03)  &    0.31 (0.06)  \\
  30 &    60.0 &    63.6 &    0.277 (0.003)  &  --0.018 (0.001)  &   0.40 (0.01)  &    0.07 (0.01)  \\
  31 &    61.3 &   128.0 &  --0.534 (0.003)  &    0.534 (0.006)  &   0.50 (0.02)  &  --0.31 (0.02)  \\
  32 &    62.9 &   918.1 &    0.000 (0.001)  &    0.000 (0.001)  &   0.00 (0.01)  &    0.00 (0.01)  \\
  33 &    63.2 &    71.9 &  --0.905 (0.002)  &  --1.579 (0.004)  &   0.35 (0.01)  &  --0.19 (0.01)  \\
 \hline
 (u)~\\
  34 &  --25.7 &    78.5 &  --2.054 (0.105)  &  --8.488 (0.119)  & --0.14 (0.22)  &  --0.27 (0.25)  \\
  35 &    40.3 &    15.3 &  --2.767 (0.011)  &  --7.023 (0.010)  & --0.16 (0.03)  &  --3.02 (0.03)  \\
  36 &    42.7 &     1.7 &  --2.765 (0.015)  &  --7.022 (0.010)  & --0.11 (0.04)  &  --2.88 (0.03)  \\
  37 &   111.9 &   107.2 &  --2.787 (0.019)  &  --6.876 (0.023)  & --0.78 (0.05)  &    1.62 (0.05)  \\
\enddata
\tablecomments {\hho maser features in \wsrc\ listed with relative
 positions $x_0$, $y_0$ and proper motions $\mu_x$, $\mu_y$ with
 respect to feature 32.
$x$ ad $y$ denote eastward and northward directions, respectively.
A proper motion of 1 \masy\ corresponds to 25.6 \kms\ at a
 distance of 5.41 kpc.
The uncertainties were obtained from the formal fitting
 uncertainties and are shown in parentheses. 
The positional offsets of $x_0$, $y_0$ relative to feature 32
 are those fitted for the mid-observation year of 2009.318 (\ie
 average year of all four epochs).
The brightness ($T_b$) of each maser feature gives the maximum
 value of all spots of the feature from all four epochs.
}\label{table:intmot}
\end{deluxetable}

\begin{deluxetable}{lllll}
\tablecolumns{5} \tablewidth{0pc} 
\tablecaption{Parallax and Proper Motion Fits for \wsrc}
\tablehead {
  \colhead{Maser \Vlsr} & \colhead{Background} &  
  \colhead{Parallax} & \colhead{$\mu_x$} &
  \colhead{$\mu_y$} 
\\
  \colhead{(\kms)}      & \colhead{Source} & 
  \colhead{(mas)} & \colhead{(\masy)} &
  \colhead{(\masy)}
            }
\startdata
 62.9       & J1922+1530 &$0.183\pm0.006$ &$-2.56\pm0.02$ &$-5.09\pm0.03$ \\
 62.9       & J1924+1540 &$0.187\pm0.009$ &$-2.49\pm0.02$ &$-5.21\pm0.04$ \\
 62.9       & Combined   &$0.185\pm0.007$ &$-2.53\pm0.02$ &$-5.15\pm0.04$ \\
\enddata
\tablecomments{Combined fit used a single parallax parameter and
 a single proper motion in each coordinate for the maser spot
 position relative to the two background sources. 
The uncertainty for each parallax fit was obtained from the
 formal fitting uncertainty.} 
\label{table:parallax}
\end{deluxetable}

\begin{deluxetable}{lrrrrrr}
\tablecolumns{7} \tablewidth{0pc} 
\tablecaption{Fitted Motions of Proto-stars and Expansion Speed}
\tablehead {
  \colhead{Source} & \colhead{$x$} & \colhead{$y$} &
  \colhead{$\Delta v_x$} & \colhead{$\Delta v_y$} &
  \colhead{$\Vlsr$} & \colhead{$V_{\rm exp}$}\\
  \colhead{} & \colhead{($''$)} & \colhead{($''$)} &
  \colhead{(\kms)} &  \colhead{(\kms)} &
  \colhead{(\kms)} & \colhead{(\kms)} 
            }
\startdata   
 e2-NW &  $-0.7 \pm 0.6$ & $ 6.3 \pm 0.8$ & $-8 \pm 6$ & $2 \pm 5$ & $66 \pm 7$ & $ 20 \pm ~9$ \\
 e2-E  &  $ 2.0 \pm 0.4$ & $ 5.8 \pm 0.5$ & $-2 \pm 6$ & $1 \pm 9$ & $60 \pm 8$ & $120 \pm 12$ \\
 e8    &  $-1.1 \pm 0.6$ & $-0.1 \pm 0.7$ & $ 6 \pm 8$ & $1 \pm 6$ & $49 \pm 7$ & $ 21 \pm 11$ \\
\enddata
\tablecomments {($x,~y$) and ($\Delta v_x,~\Delta v_y$) are
relative to the reference feature 32.
} 
\label{table:average}
\end{deluxetable}

\begin{deluxetable}{rrrrrrrr}
\tablecolumns{8} \tablewidth{0pc} 
\tablecaption{Maser Spot Motions in \wsrc}
\tablehead {
  \colhead{ID}     & \colhead{\#}     & \colhead{\Vlsr}  &
  \colhead{$T_b$}  & \colhead{$x_0$}  & \colhead{$y_0$}  &
  \colhead{$\mu_x$}& \colhead{$\mu_y$}   
\\
  \colhead{}       & \colhead{}       & \colhead{(\kms)} &
  \colhead{(Jy/b)} & \colhead{($''$)} & \colhead{($''$)} &
  \colhead{(\masy)}& \colhead{(\masy)}   
            }
\startdata
1 & 1 & 39.0 & 1.1 &--0.879 (0.015) & 7.786 (0.046) &--0.41 (0.05) & 1.35 (0.14) \\
1 & 2 & 39.4 & 1.8 &--0.879 (0.010) & 7.786 (0.030) &--0.45 (0.05) & 1.23 (0.17) \\
2 & 1 & 44.8 & 4.2 & 0.721 (0.009) & 6.177 (0.008) & 0.15 (0.02) &--0.25 (0.02) \\
2 & 2 & 45.3 & 2.7 & 0.721 (0.026) & 6.177 (0.025) &--0.02 (0.06) &--0.10 (0.05) \\
3 & 1 & 46.5 & 9.3 & 0.700 (0.015) & 6.196 (0.003) & 0.59 (0.03) &--0.49 (0.01) \\
3 & 2 & 46.9 & 5.7 & 0.700 (0.006) & 6.196 (0.007) & 0.78 (0.01) &--0.49 (0.01) \\
4 & 1 & 51.2 & 5.6 & 0.930 (0.031) & 7.216 (0.012) &--0.34 (0.08) &--0.17 (0.03) \\
4 & 2 & 51.6 & 6.0 & 0.930 (0.036) & 7.216 (0.000) &--0.29 (0.10) &--0.14 (0.01) \\
5 & 1 & 58.7 & 106.8 &--0.732 (0.040) & 7.238 (0.031) &--0.67 (0.11) &--1.25 (0.08) \\
5 & 2 & 59.2 & 127.7 &--0.732 (0.038) & 7.238 (0.027) &--0.71 (0.09) &--1.25 (0.07) \\
5 & 3 & 59.6 & 156.3 &--0.732 (0.025) & 7.238 (0.022) &--0.59 (0.05) &--1.16 (0.05) \\
5 & 4 & 60.0 & 316.5 &--0.732 (0.016) & 7.238 (0.007) &--0.53 (0.04) &--1.11 (0.02) \\
5 & 5 & 60.4 & 537.5 &--0.732 (0.020) & 7.238 (0.007) &--0.67 (0.05) &--1.18 (0.02) \\
5 & 6 & 60.8 & 644.5 &--0.732 (0.052) & 7.238 (0.031) &--0.68 (0.12) &--1.16 (0.07) \\
5 & 7 & 61.3 & 474.7 &--0.732 (0.048) & 7.238 (0.040) &--0.44 (0.21) &--1.11 (0.13) \\
6 & 1 & 61.7 & 336.9 &--0.732 (0.006) & 7.238 (0.039) & 0.02 (0.03) &--0.94 (0.18) \\
6 & 2 & 62.1 & 221.9 &--0.732 (0.016) & 7.238 (0.040) & 0.12 (0.04) &--0.87 (0.09) \\
6 & 3 & 62.5 & 138.5 &--0.732 (0.027) & 7.238 (0.048) & 0.06 (0.06) &--0.83 (0.10) \\
7 & 1 & 71.8 & 79.3 &--0.674 (0.006) & 7.518 (0.013) &--0.58 (0.02) & 0.04 (0.08) \\
7 & 2 & 72.2 & 110.6 &--0.674 (0.006) & 7.518 (0.011) &--0.54 (0.01) &--0.01 (0.04) \\
7 & 3 & 72.6 & 114.9 &--0.674 (0.006) & 7.518 (0.007) &--0.53 (0.03) &--0.02 (0.02) \\
7 & 4 & 73.1 & 91.1 &--0.674 (0.007) & 7.518 (0.008) &--0.51 (0.04) &--0.03 (0.03) \\
7 & 5 & 73.5 & 48.5 &--0.674 (0.006) & 7.518 (0.008) &--0.49 (0.02) &--0.07 (0.03) \\
7 & 6 & 73.9 & 17.8 &--0.674 (0.013) & 7.518 (0.007) &--0.52 (0.03) &--0.06 (0.02) \\
8 & 1 & 77.3 & 2.1 &--0.674 (0.067) & 7.315 (0.036) &--0.66 (0.18) &--0.07 (0.09) \\
8 & 2 & 77.7 & 2.5 &--0.674 (0.059) & 7.315 (0.030) &--0.94 (0.13) &--0.17 (0.07) \\
9 & 1 & 82.7 & 6.8 &--0.662 (0.035) & 7.416 (0.054) &--0.39 (0.22) & 0.16 (0.31) \\
9 & 2 & 83.2 & 8.8 &--0.662 (0.043) & 7.416 (0.041) &--0.31 (0.23) & 0.16 (0.21) \\
9 & 3 & 83.6 & 13.5 &--0.662 (0.048) & 7.416 (0.040) &--0.44 (0.16) & 0.14 (0.15) \\
9 & 4 & 84.0 & 13.6 &--0.662 (0.034) & 7.416 (0.035) &--0.45 (0.14) & 0.11 (0.17) \\
10 & 1 & 94.5 & 93.8 &--0.655 (0.010) & 7.367 (0.018) &--0.90 (0.03) & 1.43 (0.05) \\
10 & 2 & 94.9 & 205.1 &--0.655 (0.010) & 7.367 (0.012) &--0.88 (0.04) & 1.40 (0.05) \\
10 & 3 & 95.4 & 357.1 &--0.655 (0.012) & 7.367 (0.008) &--0.86 (0.03) & 1.41 (0.03) \\
10 & 4 & 95.8 & 357.5 &--0.655 (0.005) & 7.367 (0.005) &--0.85 (0.01) & 1.41 (0.01) \\
10 & 5 & 96.2 & 182.4 &--0.656 (0.006) & 7.367 (0.003) &--0.98 (0.02) & 1.36 (0.01) \\
10 & 6 & 96.6 & 51.7 &--0.656 (0.010) & 7.367 (0.010) &--1.06 (0.03) & 1.26 (0.08) \\
11 & 1 & 100.0 & 601.3 &--0.656 (0.024) & 7.361 (0.005) &--0.18 (0.05) & 1.28 (0.01) \\
11 & 2 & 100.4 & 280.6 &--0.656 (0.021) & 7.361 (0.005) &--0.17 (0.05) & 1.30 (0.01) \\
11 & 3 & 100.8 & 76.9 &--0.656 (0.010) & 7.361 (0.010) &--0.21 (0.02) & 1.33 (0.02) \\
11 & 4 & 101.3 & 20.1 &--0.656 (0.008) & 7.361 (0.014) &--0.22 (0.02) & 1.36 (0.03) \\
12 & 1 & 105.9 & 29.7 &--0.702 (0.024) & 7.306 (0.044) &--0.56 (0.11) & 1.22 (0.21) \\
12 & 2 & 106.3 & 48.4 &--0.702 (0.036) & 7.306 (0.034) &--0.52 (0.12) & 1.12 (0.26) \\
12 & 3 & 106.7 & 63.0 &--0.702 (0.017) & 7.306 (0.031) &--0.43 (0.13) & 1.00 (0.28) \\
12 & 4 & 107.2 & 73.4 &--0.702 (0.009) & 7.306 (0.016) &--0.50 (0.05) & 1.07 (0.20) \\
13 & 1 & 108.8 & 52.1 &--0.701 (0.082) & 7.307 (0.134) & 1.03 (0.17) & 3.17 (0.27) \\
13 & 2 & 109.7 & 103.2 &--0.701 (0.092) & 7.307 (0.114) & 0.97 (0.19) & 3.07 (0.24) \\
14 & 1 &--52.2 & 1.9 & 0.800 (0.017) & 6.033 (0.044) & 5.64 (0.04) &--3.74 (0.09) \\
14 & 2 &--51.8 & 3.6 & 0.800 (0.019) & 6.033 (0.061) & 5.65 (0.04) &--3.74 (0.14) \\
14 & 3 &--51.4 & 3.3 & 0.800 (0.026) & 6.033 (0.056) & 5.63 (0.06) &--3.75 (0.12) \\
15 & 1 &--35.4 & 1.1 & 0.706 (0.016) & 6.059 (0.017) & 5.06 (0.06) &--2.83 (0.06) \\
15 & 2 &--34.9 & 2.1 & 0.706 (0.043) & 6.059 (0.010) & 4.97 (0.12) &--2.87 (0.03) \\
15 & 3 &--34.5 & 6.1 & 0.706 (0.041) & 6.059 (0.014) & 4.97 (0.10) &--2.95 (0.04) \\
15 & 4 &--34.1 & 16.8 & 0.706 (0.047) & 6.059 (0.021) & 5.05 (0.13) &--2.95 (0.05) \\
15 & 5 &--33.7 & 18.1 & 0.706 (0.046) & 6.059 (0.019) & 4.92 (0.13) &--2.90 (0.05) \\
16 & 1 &--24.8 & 10.1 & 0.687 (0.010) & 6.095 (0.016) & 5.38 (0.08) &--3.40 (0.10) \\
16 & 2 &--24.4 & 22.3 & 0.687 (0.004) & 6.095 (0.053) & 5.31 (0.04) &--3.26 (0.35) \\
16 & 3 &--24.0 & 23.2 & 0.687 (0.015) & 6.095 (0.077) & 5.23 (0.05) &--2.98 (0.33) \\
16 & 4 &--23.6 & 30.9 & 0.687 (0.010) & 6.095 (0.005) & 5.14 (0.03) &--2.97 (0.01) \\
16 & 5 &--23.2 & 34.2 & 0.687 (0.018) & 6.095 (0.070) & 5.07 (0.05) &--3.51 (0.17) \\
16 & 6 &--22.7 & 16.6 & 0.687 (0.024) & 6.095 (0.071) & 5.06 (0.13) &--3.42 (0.28) \\
17 & 1 & 137.7 & 3.9 &--0.210 (0.036) & 6.683 (0.003) &--3.22 (0.09) & 2.26 (0.01) \\
17 & 2 & 138.1 & 4.5 &--0.210 (0.016) & 6.684 (0.015) &--3.22 (0.05) & 2.25 (0.04) \\
18 & 1 & 138.1 & 1.2 & 0.269 (0.014) & 6.323 (0.031) &--5.23 (0.05) & 1.45 (0.08) \\
18 & 2 & 138.5 & 2.7 & 0.269 (0.010) & 6.323 (0.011) &--5.25 (0.02) & 1.41 (0.02) \\
19 & 1 & 139.4 & 1.0 &--0.229 (0.016) & 6.670 (0.006) &--3.50 (0.05) & 2.17 (0.02) \\
19 & 2 & 139.8 & 0.9 &--0.229 (0.009) & 6.670 (0.007) &--3.54 (0.03) & 2.21 (0.02) \\
20 & 1 & 139.4 & 9.5 & 0.269 (0.048) & 6.323 (0.035) &--4.76 (0.10) & 1.98 (0.07) \\
20 & 2 & 139.8 & 12.0 & 0.269 (0.028) & 6.323 (0.021) &--4.67 (0.06) & 2.00 (0.05) \\
20 & 3 & 140.2 & 9.4 & 0.269 (0.031) & 6.323 (0.022) &--4.63 (0.08) & 2.03 (0.05) \\
20 & 4 & 140.6 & 3.7 & 0.269 (0.028) & 6.323 (0.049) &--4.62 (0.06) & 2.06 (0.11) \\
21 & 1 & 149.5 & 5.5 &--0.646 (0.014) & 7.125 (0.011) &--2.78 (0.04) & 3.00 (0.04) \\
21 & 2 & 149.9 & 6.2 &--0.646 (0.010) & 7.125 (0.026) &--2.83 (0.05) & 2.90 (0.15) \\
21 & 3 & 150.3 & 3.2 &--0.646 (0.017) & 7.125 (0.048) &--2.92 (0.06) & 2.73 (0.16) \\
22 & 1 & 150.7 & 3.5 &--0.649 (0.036) & 7.083 (0.037) &--2.66 (0.09) & 2.73 (0.10) \\
22 & 2 & 151.2 & 2.3 &--0.649 (0.020) & 7.083 (0.018) &--2.62 (0.06) & 2.57 (0.06) \\
23 & 1 &--29.1 & 8.1 & 0.591 (0.003) &--0.311 (0.008) & 6.77 (0.01) &--3.50 (0.02) \\
23 & 2 &--28.6 & 10.0 & 0.591 (0.003) &--0.311 (0.009) & 6.78 (0.01) &--3.50 (0.03) \\
23 & 3 &--28.2 & 5.6 & 0.591 (0.002) &--0.311 (0.013) & 6.78 (0.01) &--3.48 (0.04) \\
23 & 4 &--27.8 & 1.6 & 0.591 (0.010) &--0.311 (0.019) & 6.79 (0.04) &--3.47 (0.07) \\
24 & 1 &--8.0 & 3.8 &--0.938 (0.004) & 0.034 (0.025) &--3.82 (0.01) & 2.27 (0.07) \\
24 & 2 &--7.6 & 5.5 &--0.938 (0.007) & 0.034 (0.017) &--3.85 (0.02) & 2.19 (0.07) \\
24 & 3 &--7.2 & 4.3 &--0.938 (0.003) & 0.034 (0.018) &--3.93 (0.01) & 2.20 (0.04) \\
24 & 4 &--6.3 & 2.7 &--0.938 (0.032) & 0.034 (0.019) &--3.85 (0.07) & 2.34 (0.04) \\
25 & 1 & 26.3 & 4.5 &--0.198 (0.008) & 0.049 (0.009) & 1.84 (0.02) & 0.25 (0.02) \\
25 & 2 & 26.7 & 6.7 &--0.198 (0.006) & 0.049 (0.012) & 1.82 (0.05) & 0.23 (0.07) \\
26 & 1 & 41.1 & 54.6 & 0.365 (0.118) &--0.074 (0.099) & 3.00 (0.26) &--0.83 (0.21) \\
26 & 2 & 41.5 & 81.7 & 0.365 (0.094) &--0.074 (0.115) & 3.01 (0.19) &--0.76 (0.24) \\
27 & 1 & 42.3 & 16.3 &--1.404 (0.003) & 0.039 (0.017) &--2.81 (0.01) & 0.67 (0.04) \\
27 & 2 & 42.7 & 37.3 &--1.404 (0.005) & 0.039 (0.023) &--2.83 (0.01) & 0.64 (0.06) \\
27 & 3 & 43.2 & 36.9 &--1.404 (0.018) & 0.039 (0.022) &--2.85 (0.05) & 0.62 (0.06) \\
27 & 4 & 43.6 & 16.8 &--1.404 (0.044) & 0.039 (0.014) &--2.86 (0.12) & 0.64 (0.07) \\
27 & 5 & 44.0 & 15.5 &--1.404 (0.029) & 0.039 (0.009) &--2.82 (0.19) & 0.66 (0.04) \\
27 & 6 & 44.4 & 8.1 &--1.404 (0.023) & 0.039 (0.009) &--2.90 (0.13) & 0.68 (0.05) \\
28 & 1 & 42.3 & 11.5 & 0.365 (0.041) &--0.074 (0.031) & 1.98 (0.11) & 0.16 (0.09) \\
28 & 2 & 42.7 & 2.4 & 0.365 (0.035) &--0.074 (0.027) & 1.90 (0.09) & 0.11 (0.07) \\
29 & 1 & 52.0 & 27.8 & 0.172 (0.018) &--0.040 (0.029) & 1.48 (0.04) & 0.34 (0.07) \\
29 & 2 & 52.4 & 15.5 & 0.172 (0.018) &--0.040 (0.039) & 1.57 (0.04) & 0.26 (0.09) \\
30 & 1 & 59.6 & 37.7 & 0.277 (0.005) &--0.018 (0.004) & 0.31 (0.02) & 0.13 (0.01) \\
30 & 2 & 60.0 & 63.6 & 0.277 (0.007) &--0.018 (0.002) & 0.39 (0.03) & 0.06 (0.01) \\
30 & 3 & 60.4 & 44.8 & 0.278 (0.005) &--0.018 (0.004) & 0.52 (0.02) &--0.06 (0.02) \\
31 & 1 & 60.8 & 51.7 &--0.534 (0.005) & 0.534 (0.009) & 0.54 (0.03) &--0.37 (0.04) \\
31 & 2 & 61.3 & 128.0 &--0.534 (0.005) & 0.534 (0.009) & 0.49 (0.03) &--0.29 (0.03) \\
31 & 3 & 61.7 & 81.2 &--0.534 (0.020) & 0.534 (0.032) & 0.45 (0.05) &--0.20 (0.07) \\
32 & 1 & 62.1 & 129.7 & 0.000 (0.001) & 0.000 (0.001) & 0.01 (0.01) & 0.01 (0.01) \\
32 & 2 & 62.5 & 622.5 & 0.000 (0.001) & 0.000 (0.001) & 0.01 (0.01) &--0.01 (0.01) \\
32 & 3 & 62.9 & 918.1 & 0.000 (0.000) & 0.000 (0.000) & 0.00 (0.01) & 0.00 (0.01) \\
32 & 4 & 63.4 & 520.8 & 0.000 (0.001) & 0.000 (0.001) &--0.01 (0.01) & 0.01 (0.01) \\
33 & 1 & 62.5 & 40.7 &--0.905 (0.002) &--1.579 (0.006) & 0.34 (0.01) &--0.16 (0.01) \\
33 & 2 & 62.9 & 60.9 &--0.905 (0.004) &--1.579 (0.010) & 0.38 (0.01) &--0.23 (0.02) \\
33 & 3 & 63.4 & 71.9 &--0.905 (0.009) &--1.579 (0.013) & 0.40 (0.02) &--0.27 (0.03) \\
33 & 4 & 63.8 & 52.7 &--0.905 (0.014) &--1.579 (0.021) & 0.38 (0.04) &--0.28 (0.06) \\
34 & 1 &--28.2 & 3.6 &--2.054 (0.198) &--8.488 (0.243) &--0.17 (0.41) &--0.25 (0.50) \\
34 & 2 &--27.8 & 5.5 &--2.054 (0.163) &--8.488 (0.208) &--0.04 (0.34) &--0.07 (0.43) \\
34 & 3 &--25.7 & 78.5 &--2.055 (0.193) &--8.489 (0.182) &--0.26 (0.42) &--0.43 (0.38) \\
35 & 1 & 39.8 & 6.0 &--2.766 (0.026) &--7.023 (0.023) & 0.11 (0.09) &--2.81 (0.10) \\
35 & 2 & 40.2 & 15.3 &--2.767 (0.022) &--7.023 (0.022) &--0.06 (0.07) &--2.96 (0.05) \\
35 & 3 & 40.6 & 9.9 &--2.767 (0.015) &--7.023 (0.012) &--0.22 (0.04) &--3.09 (0.04) \\
36 & 1 & 42.3 & 0.7 &--2.765 (0.017) &--7.022 (0.015) &--0.07 (0.05) &--2.91 (0.05) \\
36 & 2 & 42.7 & 1.7 &--2.765 (0.027) &--7.022 (0.012) &--0.22 (0.09) &--2.86 (0.04) \\
37 & 1 & 111.8 & 107.2 &--2.787 (0.034) &--6.876 (0.049) &--0.70 (0.10) & 1.61 (0.16) \\
37 & 2 & 112.2 & 59.5 &--2.787 (0.045) &--6.876 (0.051) &--0.71 (0.11) & 1.56 (0.12) \\
37 & 3 & 112.6 & 16.8 &--2.787 (0.041) &--6.876 (0.049) &--0.76 (0.09) & 1.57 (0.10) \\
37 & 4 & 113.0 & 6.1 &--2.787 (0.037) &--6.876 (0.037) &--0.88 (0.08) & 1.67 (0.08) \\
\enddata
\tablecomments {
Individual \hho maser spots in \wsrc\ that are grouped as features in Table \ref{table:intmot}.
Column 1 indicates maser feature IDs corresponding to Table \ref{table:intmot}.
Column 2 shows maser spot IDs in different velocity channels in each feature.} 
\label{table:spotmot}
\end{deluxetable}

\end{document}